\documentclass[english,11pt,aps,prd,a4paper,preprintnumbers,floatfix,nofootinbib,showpacs,superscriptaddress, notitlepage]{revtex4-1} 
 \pdfoutput=1
\usepackage[usenames,dvipsnames]{color}  
\usepackage{graphicx}
\usepackage{pifont}
\usepackage{caption}
\usepackage{subcaption}
\usepackage{amsmath}
\usepackage{amssymb}
\usepackage[colorlinks=true,citecolor=darkred,urlcolor=darkred, pdfborder={0 0 0}]{hyperref}
\usepackage[normalem]{ulem}

\usepackage[T1]{fontenc}
\usepackage[latin9]{inputenc}
\usepackage{array}
\usepackage{booktabs}
\usepackage{mathrsfs}
\usepackage{multirow}
\usepackage{tabularx}
\usepackage[export]{adjustbox}
\usepackage{float}
\usepackage{multirow}

\usepackage{xcolor}

\definecolor{darkred}{rgb}{0.6,0,0}
\definecolor{dbrown}{rgb}{0.4,0.26,0.13}

\definecolor{linkcolor}{rgb}{0,0,0.5}

\def\gsim{\raise0.3ex\hbox{$\;>$\kern-0.75em\raise-1.1ex\hbox{$\sim\;$}}}
\def\lsim{\raise0.3ex\hbox{$\;<$\kern-0.75em\raise-1.1ex\hbox{$\sim\;$}}}

\def\beqn#1{\begin{equation}\label{#1}}
\def\eeqn{\end{equation}}

\def\beqa#1{\begin{eqnarray}\label{#1}}
\def\eeqa{\end{eqnarray}}

\def\Z2{$\mathcal{Z_2}$}

\newcommand {\ignore}[1]{}

\def\cevns{CE$\nu$NS}
\def\eves{E$\nu$ES}
\def\321{$\mathrm{SU(3) \otimes SU(2) \otimes U(1)}$ }
\def\d{\mathrm{d}}

\newcommand{\AddrAHEP}{%
  AHEP Group, Institut de F\'{i}sica Corpuscular --
  CSIC/Universitat de Val\`{e}ncia \\ 
 C/ Catedr\'atico Jos\'e Beltr\'an, 2 E-46980 Paterna, Spain}

\newcommand{\AddrFisteo}{%
Departament de F\'{i}sica Te\'orica, Universitat de Val\`encia, Burjassot 46100, Spain}

\newcommand{\AddrMiranda}{%
Departamento de F\'{\i}sica, Centro de Investigaci\'on
  y de Estudios Avanzados del IPN,\\ Apartado Postal 14-740 07000 Mexico,
  Distrito Federal, Mexico}

\newcommand{\AddrIoannina}{%
Division of Theoretical Physics, University of  Ioannina, GR 45110 Ioannina, Greece}

\begin{document}

\bibliographystyle{unsrt}   

\title{\boldmath \color{BrickRed} Low-energy probes of sterile neutrino transition magnetic moments} 

\author{O. G. Miranda}\email{omr@fis.cinvestav.mx}\affiliation{\AddrMiranda}

\author{D. K. Papoulias}\email{d.papoulias@uoi.gr}\affiliation{\AddrIoannina}

\author{O. Sanders}\email{osanders@fis.cinvestav.mx}
\affiliation{\AddrMiranda}

\author{M. T\'ortola}\email{mariam@ific.uv.es}
\affiliation{\AddrFisteo}\affiliation{\AddrAHEP}

\author{J. W. F. Valle}\email{valle@ific.uv.es}
\affiliation{\AddrAHEP}

\begin{abstract}
\vskip .5cm  

Sterile neutrinos with keV--MeV masses and non-zero transition magnetic moments can be probed through low-energy nuclear or electron recoil measurements. 
Here we determine the sensitivities of current and future searches, showing how they can probe a previously unexplored parameter region.
Future coherent elastic neutrino-nucleus scattering (CE$\nu$NS) or elastic neutrino-electron scattering (E$\nu$ES) experiments
using a monochromatic $^{51}$Cr source can fully probe the region indicated by the recent XENON1T excess.

\end{abstract}

\maketitle
\section{Introduction}

Despite their poor theoretical motivation, the existence of light electroweak singlet sterile neutrinos may point
to a radical departure from the Standard Model (SM), such as extra space-time dimensions~\cite{Antoniadis:1998ig,Ioannisian:1999sw,Chen:2015jta},
the existence of new protecting symmetries, such as a nearly conserved lepton number ~\cite{Peltoniemi:1992ss,Peltoniemi:1993ec},
or the presence of a singular seesaw mechanism ~\cite{Allen:1991zc}, amongst other possibilities.
From this viewpoint, probing the existence of light sterile neutrinos is extremely interesting.
Depending on the mass and mixing parameters characterizing the light sterile states, there can be stringent restrictions following from
laboratory, astrophysical and cosmological observations~\cite{Drewes:2016upu,Giunti:2019aiy,Dasgupta:2021ies,ArgoNeuT:2021clc}.

An attractive theoretical benchmark employing the type-I seesaw mechanism based upon the SM gauge symmetry~\cite{Schechter:1980gr} assumes the existence of heavy singlet ``right-handed'' neutrinos.
The existence of such sterile states at the keV--MeV mass range~\cite{Canetti:2012kh} would provide a very interesting cosmological scenario
in which the oscillations of two heavier singlet right-handed neutrinos of the seesaw mechanism produce lepton number asymmetries.
These, in turn, induce a successful leptogenesis picture that account for the observed baryon asymmetry of the Universe~\cite{Akhmedov:1998qx},
while the light right-handed neutrino accounts for cosmological dark matter~\cite{Asaka:2005pn}.
Recently, interesting variant scenarios have been discussed, see, e.g.~\cite{Drewes:2021nqr}. \\[-.4cm]

In this paper, we examine the electromagnetic properties of Majorana neutrinos~\cite{Schechter:1981hw,Li:1981um,Nieves:1981zt,Kayser:1982br,Shrock:1982sc}.
These would imply, for example, the existence of a solar antineutrino flux.
Analyzing solar neutrino data from the KamLAND experiment, one gets constraints on the Majorana neutrino transition magnetic moments (TMMs)
and solar magnetic fields~\cite{Miranda:2003yh,Miranda:2004nz}.
Likewise, it could account~\cite{Miranda:2020kwy} for the excess of electronic recoil events observed in XENON1T~\cite{Aprile:2020tmw}.
Here we focus on transition magnetic moments in the presence of light electroweak singlet neutrinos.
As a motivation, these would have an impact on experiments such as XENON1T~\cite{Brdar:2020quo,Shoemaker:2020kji,Karmakar:2020rbi} and IceCube~\cite{Coloma:2017ppo}.
Neutrino experiments looking for coherent elastic neutrino-nucleus scattering (CE$\nu$NS) or elastic neutrino-electron scattering (E$\nu$ES) events
have been proven to be a valuable tool for investigating neutrino oscillations beyond the standard  three-neutrino picture,  such as  deviations from lepton unitarity~\cite{Escrihuela:2015wra,Miranda:2020syh,Forero:2021azc},
or  the presence of light sterile neutrinos~\cite{Kosmas:2017zbh}.
Extending our previous  works, here we  examine various ways to probe sterile neutrino TMMs through nuclear and electron recoils in various experiments.\\[-.4cm]

For definiteness and simplicity, we assume just one light sterile neutrino. 
In this case, the neutrino mass matrix and the transition magnetic moments are described by 4$\times$4 symmetric and anti-symmetric matrices, respectively.
The basic \cevns\ and neutrino-electron scattering cross-sections in the SM are taken from~\cite{McKeen:2010rx,Magill:2018jla}.
We discuss current experiments such as COHERENT~\cite{Akimov:2017ade, Akimov:2020pdx}, TEXONO~\cite{Deniz:2009mu} and XENON1T~\cite{Aprile:2020tmw},
as well as the potential of future \cevns\ and E$\nu$ES experiments using a $^{51}$Cr monochromatic neutrino source~\cite{Bellenghi:2019vtc,Link:2019pbm}.

The paper is organized as follows. The basic formalism describing neutrino magnetic moments in the presence of massive singlet leptons is given in Sec.~\ref{sec:basic-formalism}.
Section \ref{sec:experiments} describes various experimental ways of probing neutrino magnetic moments, several of which are new proposals.
Our estimated experimental sensitivities are presented in Sec.~\ref{sec:results}, and a final summary and outlook is provided in Sec.~\ref{sec:conclusions-outlook}.

\section{Basic formalism}
\label{sec:basic-formalism}

\subsection{ Neutrino magnetic moment and massive ``sterile'' leptons}
\label{sec:neutr-magn-moment}

For the case of Majorana particles, the electromagnetic interaction Hamiltonian in the mass basis is given by~\cite{Schechter:1981hw} 
 \begin{equation}
\textit{H}^{M}_{em} = -\frac{1}{4}\nu^{T}_{L}C^{-1}~ \tilde{\lambda}~\sigma^{\alpha\beta}\nu_{L}F_{\alpha\beta} + h.c. ,
\end{equation}
where $\tilde{\lambda}$ is the complex antisymmetric matrix describing transition moments, and contains the information on the magnetic and electric dipole moments, $\tilde{\lambda} = \mu - id$.

  In this work, we will assume that neutrinos are Majorana fermions, and we will encode the magnetic and dipole moment information in the aforementioned parameter, $\tilde{\lambda}$.
  We will also assume that, in addition to the three light active neutrino states, we have extra mass eigenstates associated to gauge singlets.
  For definiteness, we assume only one of such ``sterile'' states, so the matrices describing the electromagnetic interaction will be 4$\times$4,
\begin{equation}
\\ \tilde{\lambda} =
\left(\begin{array}{cccc}
  0 & \tilde{\lambda}_{12} & \tilde{\lambda}_{13} & \tilde{\lambda}_{14} \\
  -\tilde{\lambda}_{12} & 0 & \tilde{\lambda}_{23} & \tilde{\lambda}_{24}  \\
  -\tilde{\lambda}_{13} & -\tilde{\lambda}_{23} & 0 & \tilde{\lambda}_{34}  \\
  -\tilde{\lambda}_{14} & -\tilde{\lambda}_{24} & - \tilde{\lambda}_{34} & 0 \end{array}\right) \, .
\label{Eq:lambda-nmm}
\end{equation}
Electromagnetic interactions between the (mainly) active to the (mainly) sterile massive states are described by the last row (column) of the matrix,
while the flavor transition moments correspond to the  upper-left 3$\times$3
sub-block~\footnote{As an academic note, we mention that in the (now unphysical) limit where the ``active'' and ``sterile'' masses coincide,
  this interaction would correspond to the often discussed ``Dirac'' neutrino magnetic moment.}.
Note that the entries $\tilde{\lambda}_{ij}$ of the transition magnetic moment matrix shown in Eq.~(\ref{Eq:lambda-nmm}) are complex numbers,
parametrized in terms of their moduli and the associated CP-violating phases
\begin{equation}
  \label{eq:lambdas}
  \tilde{\lambda}_{ij} = \vert \tilde{\lambda}_{ij}\vert e^{i \zeta_{ij}}. 
\end{equation}

We now turn to the determination of the expressions for the effective neutrino magnetic moment associated to the experimental setup of interest.
We start with the case of solar neutrino experiments,  generalizing previous results~\cite{Vogel:1989iv,Beacom:1999wx} to a general expression in the mass basis  
\begin{equation}
\left(\mu^M_{\nu,\text{eff}}\right)^2 (L, E_\nu) = \sum_j \Big \vert \sum_i K^\ast_{\alpha i} e^{-i\, \Delta m^2_{ij} L /2 E_\nu} \tilde{\lambda}_{ij} \Big \vert^2 \, ,
\label{NMM-eff}
\end{equation}
where the 3$\times$(3+m) rectangular matrix $K$ is the upper truncation of the (3+m)$\times$(3+m) unitary matrix diagonalizing the neutrinos, where m is the number of sterile neutrinos~\cite{Schechter:1980gr}.
We also assume that the charged leptons are in their mass-diagonal basis.
The indices  $i$ and $j$ run over the total number of neutrino mass eigenstates.  
The large baseline-distance for the case of solar neutrinos makes the interference terms vanish.
Therefore, we can generalize the three-neutrino expression in Ref.~\cite{Grimus:2002vb} to our case of interest, where we take just four neutrino species (m=1), 
\begin{eqnarray}\label{eq:nmm_solar}
(\mu^{M}_{\nu,\,\text{sol}})^{2} &=& 
  P_{e1}(|\tilde{\lambda}_{12}|^{2} + |\tilde{\lambda}_{13}|^{2} + |\tilde{\lambda}_{14}|^{2})  +
  P_{e2}(|\tilde{\lambda}_{12}|^{2} + |\tilde{\lambda}_{23}|^{2} + |\tilde{\lambda}_{24}|^{2})  \\ &+& 
  P_{e3}(|\tilde{\lambda}_{13}|^{2} + |\tilde{\lambda}_{23}|^{2} + |\tilde{\lambda}_{34}|^{2}) +
    P_{e4}(|\tilde{\lambda}_{14}|^{2} + |\tilde{\lambda}_{24}|^{2} + |\tilde{\lambda}_{34}|^{2}) \, , \nonumber 
\end{eqnarray}
where $P_{ei}$ corresponds to the solar neutrino transition probability from the originally created $\nu_e$ state to the mass eigenstate $\nu_i$ (see Ref.~\cite{Miranda:2020kwy}).   

For short-baseline experiments, and assuming very small mixing between active and sterile states, one can just assume at detectors the presence of pure $\nu_e$ and $\nu_\mu$ beams.
In this case, the full analytical expressions for the effective neutrino magnetic moments for $\nu_e$ and $\nu_\mu$ in the mass basis are lengthy.
For the reader's convenience, we give them in the form 
\begin{equation}
\begin{aligned}
\left(\mu^{M}_{\nu_e}\right)^2 &= \sum_{\kappa} \tilde{\Lambda}^2_\kappa \, C_{\nu_e}^{\kappa}(\theta_{ij}, \zeta_{ij}, \delta_{ij})\, , \\
\left(\mu^{M}_{\nu_\mu}\right)^2 &= \sum_{\kappa} \tilde{\Lambda}^2_\kappa \, C_{\nu_\mu}^{\kappa}(\theta_{ij}, \zeta_{ij}, \delta_{ij})\, ,
\end{aligned}
\label{eq:munue_massbasis}
\end{equation} 
where the index $\kappa$ runs over the 21 components listed in Table~\ref{tab:coefficients_nue} (for $\alpha={e}$) and Table~\ref{tab:coefficients_numu} (for $\alpha={\mu}$). 
Here, $\tilde{\Lambda}_k^2$ has dimensions of $\mu_B^2$ and represents the product of two TMMs (all possible combinations of $\vert \tilde{\lambda}_{ij} \vert \vert \tilde{\lambda}_{i^\prime j^\prime} \vert $),
and $C_{\nu_\alpha}^{\kappa}(\theta_{ij}, \zeta_{ij}, \delta_{ij})$ denotes the associated coefficients.
At this point, we should stress that the expressions in Eq.~(\ref{eq:munue_massbasis}) remain the same for the case of antineutrinos.
Notice that, when the sterile neutrino parameters are neglected, Eq.~(\ref{eq:munue_massbasis}) reduces to the three-neutrino mixing expression obtained in Ref.~\cite{Miranda:2019wdy}. 
Since we are interested in TMMs from active to sterile states, in the following calculations we will concentrate only on the relevant terms containing $\tilde{\lambda}_{i4}$.
Thus, ignoring the active-active terms with $\tilde{\lambda}_{12}$, $\tilde{\lambda}_{13}$ and $\tilde{\lambda}_{23}$ as well as the cross terms $\tilde{\lambda}_{i4} \tilde{\lambda}_{j4}$,
the effective magnetic moment responsible for $\nu_e \to \nu_s$ and $\bar{\nu}_e \to \nu_s$ transitions reduces to 
\begin{equation}
\begin{aligned}
\left(\mu_{\nu_{e}\to \nu_s}^{M}\right)^2 = \vert \tilde{\lambda}_{14}\vert^2 \left(c_{12}^2 c_{13}^2 c_{14}^2+s_{14}^2\right)+ \vert\tilde{\lambda} _{24}\vert^2 \left(s_{12}^2 c_{13}^2 c_{14}^2 +s_{14}^2\right) +   \vert\tilde{\lambda}_{34}\vert^2 \left(  s_{13}^2 c_{14}^2+s_{14}^2 \right) \, .
\end{aligned}
\end{equation}
Here, $c_{ij} = \cos\theta_{ij}$ and $s_{ij} = \sin\theta_{ij}$, with $\theta_{ij}$ being the different neutrino mixing angles
and $\delta_{ij}$ denoting the corresponding CP violating phases present in the four-neutrino mixing matrix.
Similarly, for the case of $\nu_\mu \to \nu_s$ or $\bar{\nu}_\mu \to \nu_s$ transitions, we get 
\begin{equation}
\begin{aligned}
\left(\mu_{\nu_{\mu}\to \nu_s}^{M}\right)^2 = & \vert\tilde{\lambda}_{14}\vert^2 \Big[   c_{24}^2 \left( \sin 2 \theta _{12} s_{13} c_{23}  s_{23} \cos\delta + s_{12}^2 c_{23}^2 +c_{12}^2 s_{13}^2 s_{23}^2\right) \\ 
&+c_{12} c_{13} s_{14} \sin 2 \theta _{24} \left[s_{12} c_{23}  \cos\delta_{14} +c_{12} s_{13} s_{23} \cos \left(\delta -\delta _{14}\right)\right]+s_{24}^2 \left(c_{12}^2 c_{13}^2 s_{14}^2+c_{14}^2\right) \Big] \\
   + & \vert\tilde{\lambda}_{24}\vert^2 \Big[ c_{24}^2 \left(-   \sin 2 \theta _{12} s_{13} c_{23}  s_{23} \cos\delta  + c_{12}^2 c_{23}^2+s_{12}^2 s_{13}^2 s_{23}^2\right)\\&+ s_{12} c_{13}  s_{14} \sin 2 \theta _{24} \left[s_{12} s_{13} s_{23} \cos \left(\delta -\delta
   _{14}\right)-c_{12} c_{23} \cos\delta _{14} \right]+s_{24}^2 \left(s_{12}^2 c_{13}^2  s_{14}^2+c_{14}^2\right)\Big] \\
   + & \vert\tilde{\lambda}_{34}\vert^2 \Big[- \sin 2 \theta _{13} s_{23}  s_{14} c_{24} s_{24}  \cos \left(\delta -\delta _{14}\right)  + c_{13}^2 s_{23}^2 c_{24}^2 +s_{24}^2 \left(s_{13}^2 s_{14}^2 + c_{14}^2\right) \Big] \, .
\end{aligned}
\end{equation}
Assuming a small active-sterile neutrino mixing,  $\sin^2\theta_{i4}\leq 0.01$, the last two expressions simplify considerably and read 
\begin{equation}
\left(\mu_{\nu_{e}\to \nu_s}^{M}\right)^2   \approx \vert\tilde{\lambda}_{14}\vert^2 \,  c_{13}^2  c_{12}^2  + \vert\tilde{\lambda}_{24}\vert^2 \, c_{13}^2  s_{12}^2 + \vert\tilde{\lambda}_{34}\vert^2  \, s_{13}^2  \, ,
\label{eq:simple-active-sterile-nue}
\end{equation}
\begin{equation}
\begin{aligned}
\left(\mu_{\nu_{\mu}\to \nu_s}^{M}\right)^2  \approx & ~\vert\tilde{\lambda}_{14}\vert^2 \left( \,\, c_{23} s_{13} s_{23} \sin 2\theta_{12} \cos \delta +c_{23}^2 s_{12}^2+c_{12}^2 s_{13}^2 s_{23}^2\right) \\
  & +\vert\tilde{\lambda}_{24}\vert^2 \left(- c_{23} s_{13} s_{23} \sin 2\theta_{12} \cos \delta  + c_{23}^2 c_{12}^2 + s_{12}^2 s_{13}^2 s_{23}^2\right) 
    + \vert\tilde{\lambda}_{34}\vert^2 c_{13}^2  s_{23}^2 \, .
\end{aligned}
\label{eq:simple-active-sterile-numu}
\end{equation}

In what follows, for our analysis involving reactor neutrino experiments ($\bar{\nu}_e$ source) or $^{51}$Cr experiments ($\nu_e$ source),
the effective neutrino magnetic moment will be given by Eq.~(\ref{eq:simple-active-sterile-nue}).
Similarly, for the analysis of the COHERENT experiment involving pion-decay-at-rest neutrinos, Eq.~(\ref{eq:simple-active-sterile-nue}) is relevant for the $\nu_e$ component of the SNS beam
and Eq.~(\ref{eq:simple-active-sterile-numu}) for the $\nu_\mu$ or $\bar{\nu}_\mu$ components.
Before closing this discussion, we comment on the phase counting.
As shown in Ref.~\cite{Grimus:2000tq}, for the case of $n$ Majorana neutrinos the number of physical phases is $n(n-2)$,
hence we have in general 8 physical phases expected in the present study (see Tables~\ref{tab:coefficients_nue} and~\ref{tab:coefficients_numu}).
\begin{table}
\begin{tabular}{|c|c|}
  \hline
  $\tilde{\Lambda}^2_\kappa$ & $C^\kappa_{\nu_e}(\theta_{ij}, \zeta_{ij}, \delta_{ij})$\\
\hline
$ \vert \tilde{\lambda} _{12} \vert^2$ & $c_{13}^2 c_{14}^2$ \\
$\vert \tilde{\lambda} _{13} \vert^2$& $c_{14}^2 (c_{12}^2 c_{13}^2 + s_{13}^2 ) $\\
$\vert \tilde{\lambda} _{14} \vert^2$ & $c_{12}^2 c_{13}^2 c_{14}^2+s_{14}^2$ \\
$\vert \tilde{\lambda} _{23}\vert^2 $& $ c_{14}^2 (s_{12}^2 c_{13}^2 +  s_{13}^2 )$\\
$\vert \tilde{\lambda} _{24} \vert^2$ & $ s_{12}^2 c_{13}^2 c_{14}^2 +s_{14}^2 $\\
$\vert \tilde{\lambda} _{34} \vert^2$& $c_{14}^2 s_{13}^2+s_{14}^2 $\\
$\vert \tilde{\lambda} _{12} \vert \vert \tilde{\lambda} _{13} \vert$ &$- c_{14}^2 s_{12} \sin 2 \theta _{13}  \cos \left(\delta -\zeta _{12}+\zeta _{13}\right) $\\
$\vert \tilde{\lambda} _{12} \vert \vert \tilde{\lambda} _{14} \vert$& $c_{13} s_{12} \sin 2 \theta _{14} \cos \left(\delta _{14}-\zeta _{12}+\zeta _{14}\right) $\\
 $ \vert\tilde{\lambda} _{12} \vert \vert\tilde{\lambda} _{23} \vert$& $-c_{14}^2 c_{12} \sin 2 \theta _{13}  \cos \left(\delta -\zeta _{12}+\zeta _{23}\right) $\\
$\vert \tilde{\lambda} _{12} \vert \vert\tilde{\lambda} _{24} \vert$ &$ - c_{13} c_{12} \sin 2 \theta _{14} \cos \left(\delta _{14}-\zeta _{12}+\zeta _{24} \right)$ \\
$\vert \tilde{\lambda} _{12} \vert \vert\tilde{\lambda} _{34}\vert$ &$ 0 $\\
 $ \vert\tilde{\lambda} _{13} \vert \vert \tilde{\lambda} _{14} \vert$& $- s_{13} \sin 2 \theta _{14}  \cos \left(\delta -\delta _{14}+\zeta _{13}-\zeta _{14}\right) $ \\
 $ \vert\tilde{\lambda} _{13} \vert \vert\tilde{\lambda} _{23} \vert$&$ - c_{13}^2 c_{14}^2 \sin 2 \theta _{12} \cos \left(\zeta _{13}-\zeta _{23}\right) $  \\
$\vert \tilde{\lambda} _{13}\vert \vert \tilde{\lambda} _{24}\vert$ &$ 0 $\\
$\vert \tilde{\lambda} _{13}\vert \vert \tilde{\lambda} _{34} \vert$ &$ c_{12} c_{13} \sin 2 \theta _{14} \cos \left(\delta _{14}-\zeta _{13}+\zeta _{34}\right) $\\
 $ \vert \tilde{\lambda} _{14} \vert \vert\tilde{\lambda} _{23} \vert$&$ 0$ \\
 $\vert \tilde{\lambda} _{14}\vert \vert \tilde{\lambda} _{24} \vert$ & $ c_{13}^2 c_{14}^2 \sin 2 \theta _{12} \cos \left(\zeta _{14}-\zeta _{24}\right)  $\\
$\vert \tilde{\lambda} _{14} \vert \vert\tilde{\lambda} _{34} \vert$&$ c_{12} c_{14}^2 \sin 2 \theta _{13} \cos \left(\delta -\zeta _{14}+\zeta _{34}\right)$ \\
$\vert \tilde{\lambda} _{23}\vert \vert \tilde{\lambda} _{24} \vert$&$ s_{13} \sin 2 \theta _{14}  \cos \left(\delta -\delta _{14}+\zeta _{23}-\zeta _{24}\right) $\\
$\vert \tilde{\lambda} _{23}\vert \vert \tilde{\lambda} _{34} \vert$&$ - c_{13} s_{12} \sin 2 \theta _{14} \cos \left(\delta _{14}-\zeta _{23}+\zeta _{34}\right) $ \\
$ \vert\tilde{\lambda} _{24}\vert \vert\tilde{\lambda} _{34} \vert$&$ c_{14}^2 s_{12} \sin 2 \theta _{13}  \cos \left(\delta -\zeta _{24}+\zeta _{34}\right) $\\
 \hline
\end{tabular}
\caption{TMMs and the corresponding coefficients entering in the expression of the effective magnetic moment in Eq.~(\ref{eq:munue_massbasis})  for electron neutrinos.}

\label{tab:coefficients_nue}
\end{table}
\begin{table}
\begin{tabular}{|c|c|}
  \hline
  $\tilde{\Lambda}^2$ & $C_{\nu_\mu}(\theta_{ij}, \zeta_{ij}, \delta_{ij})$\\
\hline 
$\vert\tilde{\lambda}_{12}\vert^2 $& $c_{24} s_{14} s_{23} s_{24}  \sin 2 \theta _{13} \cos \left(\delta -\delta _{14}\right) + c_{24}^2 \left(c_{23}^2+s_{13}^2 s_{23}^2\right)+c_{13}^2 s_{14}^2 s_{24}^2 $\\
\hline
 $\vert\tilde{\lambda}_{13}\vert^2 $& $\begin{aligned} &c_{24}^2 \left[c_{23} s_{13} s_{23} \sin 2 \theta _{12} \cos \delta  + c_{23}^2 s_{12}^2+s_{23}^2 \left(c_{12}^2 s_{13}^2+c_{13}^2\right)\right]\\+&c_{13} s_{14} \sin 2 \theta _{24} \left[c_{12} c_{23} s_{12} \cos
   \delta _{14} +\left(c_{12}^2-1\right) s_{13} s_{23} \cos \left(\delta -\delta _{14}\right)\right] +s_{14}^2 s_{24}^2 \left(c_{12}^2 c_{13}^2+s_{13}^2\right)\end{aligned}$ \\
   \hline
$ \vert\tilde{\lambda}_{14}\vert^2$ & $ \begin{aligned} & c_{24}^2 \left[c_{23} s_{13} s_{23}  \sin 2 \theta _{12}  \cos \delta  +c_{23}^2 s_{12}^2+c_{12}^2 s_{13}^2 s_{23}^2\right]\\ &+c_{12} c_{13} s_{14} \sin 2 \theta _{24} \left[c_{23} s_{12} \cos \delta_{14}+c_{12} s_{13} s_{23} \cos \left(\delta -\delta _{14}\right)\right]+s_{24}^2 \left(c_{12}^2 c_{13}^2 s_{14}^2+c_{14}^2\right) \end{aligned}$\\
   \hline
$ \vert\tilde{\lambda}_{23}\vert^2 $& $\begin{aligned} & c_{24}^2 \left[-c_{23} s_{13} s_{23}  \sin 2 \theta _{12} \cos \delta + s_{23}^2 \left(c_{13}^2+s_{12}^2 s_{13}^2\right)+c_{12}^2 c_{23}^2\right]\\&-c_{13} s_{14} \sin 2 \theta _{24}  \left[c_{12} c_{23} s_{12} \cos \delta_{14} -\left(s_{12}^2-1\right) s_{13} s_{23} \cos \left(\delta -\delta _{14}\right)\right]+s_{14}^2 s_{24}^2 \left(c_{13}^2 s_{12}^2+s_{13}^2\right)\end{aligned}$ \\
   \hline
$ \vert\tilde{\lambda}_{24}\vert^2$ &$\begin{aligned} & c_{24}^2 \left[-c_{23} s_{13} s_{23}   \sin 2 \theta_{12}  \cos \delta + c_{12}^2 c_{23}^2+s_{12}^2 s_{13}^2 s_{23}^2\right]\\&+c_{13} s_{12} s_{14} \sin 2 \theta_{24}  \left[s_{12} s_{13} s_{23} \cos \left(\delta -\delta
   _{14}\right)-c_{12} c_{23} \cos \delta_{14}  \right]+s_{24}^2 \left(c_{13}^2 s_{12}^2 s_{14}^2+c_{14}^2\right)\end{aligned}$ \\
   \hline
$\vert \tilde{\lambda}_{34}\vert^2$ &$  -c_{24} s_{14} s_{24} s_{23} \sin 2 \theta _{13}  \cos \left(\delta -\delta _{14}\right)  +c_{13}^2 c_{24}^2 s_{23}^2+s_{24}^2 \left(c_{14}^2+s_{13}^2 s_{14}^2\right)$ \\
\hline
 $\vert\tilde{\lambda}_{12}\vert \vert \tilde{\lambda}_{13}\vert$ & $\begin{aligned} -s_{12} \Big[& - c_{13}^2 s_{14} s_{23} \sin 2 \theta_{24}  \cos \left(\delta _{14}-\zeta _{12}+\zeta _{13}\right) \\&+\left(s_{14}^2 s_{24}^2-c_{24}^2 s_{23}^2\right) \sin 2 \theta_{13} \cos \left(\delta -\zeta _{12}+\zeta _{13}\right)+s_{13}^2 s_{14} s_{23} \sin 2 \theta_{24}  \cos \left(2 \delta -\delta _{14}-\zeta _{12}+\zeta _{13}\right)\Big]\\
   &-2 c_{12} c_{23} c_{24} \left[c_{13} c_{24} s_{23} \cos \left(\zeta _{12}-\zeta _{13}\right)-s_{13} s_{14} s_{24} \cos \left(\delta -\delta _{14}-\zeta _{12}+\zeta _{13}\right)\right] \end{aligned}$ \\
   \hline
$ \vert\tilde{\lambda}_{12} \vert \vert \tilde{\lambda}_{14}\vert$ &$ \begin{aligned} -2 c_{14} s_{24} \left\{ s_{12} \left[c_{24} s_{13} s_{23} \cos \left(\delta -\zeta _{12}+\zeta _{14}\right)+c_{13} s_{14} s_{24} \cos \left(\delta _{14}-\zeta _{12}+\zeta _{14}\right)\right] -c_{12} c_{23} c_{24} \cos \left(\zeta_{12}-\zeta_{14}\right)\right\} \end{aligned}$\\
   \hline
$ \vert\tilde{\lambda}_{12}\vert \vert \tilde{\lambda}_{23}\vert$ &$\begin{aligned} &c_{12} c_{13}^2 s_{14} s_{23} \sin 2 \theta_{24}  \cos \left(\delta _{14}-\zeta _{12}+\zeta _{23}\right)\\&+c_{13} \left[2 c_{12} s_{13} \left(c_{24}^2 s_{23}^2-s_{14}^2 s_{24}^2\right) \cos \left(\delta -\zeta _{12}+\zeta _{23}\right)+c_{24}^2 s_{12} \cos \left(\zeta _{12}-\zeta _{23}\right) \sin 2 \theta_{23} \right]\\&-s_{13} s_{14} \sin 2 \theta_{24}  \left[c_{23} s_{12} \cos \left(\delta -\delta _{14}-\zeta _{12}+\zeta _{23}\right)+c_{12} s_{13} s_{23} \cos \left(2 \delta -\delta _{14}-\zeta _{12}+\zeta _{23}\right)\right] \end{aligned} $\\
   \hline
$ \vert \tilde{\lambda}_{12} \vert \vert \tilde{\lambda}_{24}\vert$ & $2 c_{14} s_{24} \left\{c_{12} \left[c_{24} s_{13} s_{23} \cos \left(\delta -\zeta _{12}+\zeta _{24}\right)+c_{13} s_{14} s_{24} \cos \left(\delta
   _{14}-\zeta _{12}+\zeta _{24}\right)\right]+c_{23} c_{24} s_{12} \cos \left(\zeta _{12}-\zeta _{24}\right)\right\} $\\
   \hline
$ \vert \tilde{\lambda}_{12} \vert \vert \tilde{\lambda}_{34}\vert $ & $0$ \\
\hline
 $\vert\tilde{\lambda}_{13} \vert \vert\tilde{\lambda}_{14} \vert$ & $2 c_{14} s_{24} \left[ s_{13} s_{14} s_{24} \cos \left(\delta -\delta _{14}+\zeta _{13}-\zeta _{14}\right)-c_{13} c_{24} s_{23} \cos \left(\zeta _{13}-\zeta _{14}\right)\right]$ \\
 \hline
$\vert \tilde{\lambda}_{13} \vert \vert \tilde{\lambda}_{23}\vert$ &$ \begin{aligned}   &  \left[-2 c_{13} c_{24} s_{13} s_{14} s_{23} s_{24} \cos \left(\delta -\delta _{14}\right)+c_{24}^2 \left(c_{23}^2-s_{13}^2
   s_{23}^2\right)-c_{13}^2 s_{14}^2 s_{24}^2\right] \sin 2 \theta_{12} \cos \left(\zeta _{13}-\zeta _{23}\right) \\&+2c_{23} c_{24} c_{12}^2 \left[c_{24} s_{13} s_{23} \cos \left(\delta -\zeta _{13}+\zeta _{23}\right)+c_{13} s_{14} s_{24} \cos \left(\delta _{14}-\zeta _{13}+\zeta _{23}\right)\right]\\&-2c_{23} c_{24} s_{12}^2 \left[c_{24}
   s_{13} s_{23} \cos \left(\delta +\zeta _{13}-\zeta _{23}\right)+c_{13} s_{14} s_{24} \cos \left(\delta _{14}+\zeta _{13}-\zeta _{23}\right)\right]  \end{aligned}$\\
   \hline
$ \vert\tilde{\lambda}_{13}\vert \vert \tilde{\lambda}_{24} \vert$& $0$ \\
\hline
 $ \vert\tilde{\lambda}_{13} \vert \vert \tilde{\lambda}_{34}\vert $&$ -2 c_{14} s_{24} \left\{c_{12} \left[c_{24} s_{13} s_{23} \cos \left(\delta -\zeta _{13}+\zeta _{34}\right)+c_{13} s_{14} s_{24} \cos \left(\delta _{14}-\zeta _{13}+\zeta _{34}\right)\right]+c_{23} c_{24} s_{12} \cos \left(\zeta
   _{13}-\zeta _{34}\right)\right\}$ \\
   \hline
$ \vert\tilde{\lambda}_{14} \vert \vert \tilde{\lambda}_{23}\vert$ &$ 0$ \\
\hline
$ \vert \tilde{\lambda}_{14} \vert \vert \tilde{\lambda}_{24} \vert$ &$ \begin{aligned} &  \left[2 c_{13} c_{24} s_{13} s_{14} s_{23} s_{24} \cos \left(\delta -\delta _{14}\right)+c_{24}^2 \left(s_{13}^2 s_{23}^2-c_{23}^2\right)+c_{13}^2
   s_{14}^2 s_{24}^2\right] \sin 2 \theta_{12} \cos \left(\zeta _{14}-\zeta _{24}\right) \\&
   -2 c_{23} c_{24} c_{12}^2 \left[c_{24} s_{13} s_{23} \cos \left(\delta -\zeta _{14}+\zeta _{24}\right)+c_{13} s_{14} s_{24} \cos \left(\delta _{14}-\zeta _{14}+\zeta _{24}\right)\right]\\&+2 c_{23} c_{24} s_{12}^2 \left[c_{24} s_{13} s_{23} \cos
   \left(\delta +\zeta _{14}-\zeta _{24}\right)+c_{13} s_{14} s_{24} \cos \left(\delta _{14}+\zeta _{14}-\zeta _{24}\right)\right] \end{aligned}$\\
   \hline
$ \vert\tilde{\lambda}_{14} \vert \vert \tilde{\lambda}_{34} \vert$ &$ \begin{aligned}   & - c_{12} c_{13}^2 s_{14} s_{23} \sin 2 \theta_{24}  \cos \left(\delta
   _{14}-\zeta _{14}+\zeta _{34}\right) \\
   - &c_{13} \left[2 c_{12} s_{13} \left(c_{24}^2
   s_{23}^2-s_{14}^2 s_{24}^2\right) \cos \left(\delta -\zeta _{14}+\zeta
   _{34}\right)+c_{24}^2 s_{12} \cos \left(\zeta _{14}-\zeta _{34}\right) \sin 2 \theta_{23} \right]\\
   + &s_{13} s_{14} \sin 2 \theta_{24}  \left[c_{23}
   s_{12} \cos \left(\delta -\delta _{14}-\zeta _{14}+\zeta _{34}\right)+c_{12} s_{13}
   s_{23} \cos \left(2 \delta -\delta _{14}-\zeta _{14}+\zeta _{34}\right)\right] \end{aligned}$\\
   \hline
$ \vert\tilde{\lambda}_{23} \vert \vert \tilde{\lambda}_{24} \vert$ &$ 2 c_{14} s_{24} \left[c_{13} c_{24} s_{23} \cos \left(\zeta _{23}-\zeta _{24}\right)-s_{13} s_{14} s_{24} \cos \left(\delta -\delta _{14}+\zeta _{23}-\zeta _{24}\right)\right]$ \\
\hline
$ \vert \tilde{\lambda}_{23} \vert \vert \tilde{\lambda}_{34} \vert$ & $2 c_{14} s_{24} \left\{s_{12} \left[c_{24} s_{13} s_{23} \cos \left(\delta -\zeta _{23}+\zeta _{34}\right)+c_{13} s_{14} s_{24} \cos \left(\delta _{14}-\zeta _{23}+\zeta _{34}\right)\right]-c_{12} c_{23} c_{24} \cos \left(\zeta
   _{23}-\zeta _{34}\right)\right\}$ \\
   \hline
$ \vert \tilde{\lambda}_{24} \vert \vert \tilde{\lambda}_{34} \vert$ & $ \begin{aligned}    s_{12} \Big[&- c_{13}^2 s_{14} s_{23} \sin 2 \theta_{24}  \cos \left(\delta _{14}-\zeta _{24}+\zeta _{34}\right)+ \left(s_{14}^2 s_{24}^2-c_{24}^2 s_{23}^2\right) \sin 2 \theta_{13}  \cos \left(\delta -\zeta _{24}+\zeta _{34}\right)\\&+s_{13}^2 s_{14} s_{23} \sin 2 \theta_{24}  \cos \left(2 \delta -\delta _{14}-\zeta _{24}+\zeta _{34}\right)\Big]
   \\&+2c_{12} c_{23} c_{24} \left[ c_{13} c_{24} s_{23} \cos \left(\zeta
   _{24}-\zeta _{34}\right)-s_{13} s_{14} s_{24} \cos \left(\delta -\delta _{14}-\zeta _{24}+\zeta _{34}\right)\right]   \end{aligned}$\\
 \hline
\end{tabular}
\caption{TMMs and corresponding coefficients for the calculation of Eq.~(\ref{eq:munue_massbasis}) for muon neutrinos.}
\label{tab:coefficients_numu}
\end{table}

\subsection{\cevns\ and neutrino-electron scattering in the SM}
\label{sec:cevns-neutr-electr}

Assuming only SM interactions, the differential \cevns\ cross-section for the nuclear recoil energy $E_r$, is written as~\cite{Barranco:2005yy} 
\begin{equation}
\left(\frac{\d \sigma}{\d E_r}\right)_{\text{SM}} = \frac{G_F^2 M}{\pi}  (\mathcal{Q}_V)^2 \left[ 1 - \frac{E_r}{E_\nu} - \frac{M E_r}{2 E_\nu^2}\right]  \, ,
\label{eq:xsec-cevns}
\end{equation}
where $G_F$, $E_\nu$, and $M$ stand for the Fermi constant, the incident neutrino energy, and the nuclear mass, respectively.
Here $\mathcal{Q}^V_W$ denotes the weak charge, defined as~\cite{Papoulias:2015vxa}
\begin{equation}
\mathcal{Q}_V =  \left[ \left(\frac12-2 \sin^2 \theta_W \right) Z F_{p}(Q^{2})  -\frac12 N F_{n}(Q^{2}) \right]  \, ,
\label{eq:Qw}
\end{equation}
with the weak mixing angle being $\sin^2 \theta_W  = 0.2312$. 
It is noteworthy that the main uncertainty in the theoretical \cevns\ calculation arises from nuclear physics effects, which may
limit the experimental sensitivity for searches of physics beyond the SM (for a detailed analysis, see Ref.~\cite{Papoulias:2019lfi}). 
In the latter expression, nuclear structure corrections are taken into account through the  nuclear form factors for
protons $F_p(Q^2)$ and neutrons $F_n(Q^2)$. The magnitude of the momentum transfer is $Q=\sqrt{2 M E_r}$. 
We employ the Helm parametrization~\footnote{Following Ref.~\cite{Akimov:2017ade}, for the COHERENT-CsI detector we adopt the Klein-Nystrand (KN) form factor 
$F_{\text{KN}} = 3 \frac{j_1(Q R_A)}{Q R_A} \left[ 1 + (Q a_k )^2 \right]^{-1} $, where $R_A= 1.23 \times A^{1/3}$ and $a_k = 0.7$ fm~\cite{Klein:1999qj}. } 
\begin{equation}
  F_{p,n}(Q^2) = 3\frac{j_1(QR_0)}{QR_0}\exp(-Q^2s^2/2) ,
  \end{equation}
where $j_1(x)=\sin(x)/x^2 - \cos(x)/x$ is the spherical Bessel function of order one, and $R_0^2=\frac53(R_{p,n}^2-3s^2)$ with $R_{p,n}$ the proton and neutron r.m.s. radii and $s=0.9$~fm the surface thickness.
\\[-.4cm]

Turning to elastic neutrino electron scattering, E$\nu$ES, for a given neutrino  flavor $\alpha=e,\mu, \tau$,
the corresponding SM differential cross-section as a function of the  electron recoil energy is given as
\begin{equation}
\left(\frac{\d \sigma_{\nu_\alpha-e^-}}{\d E_r}\right)_\text{SM}^\text{free} =  \frac{2 G_F^2 m_e}{\pi} \left[g_L^2 + g_R^2 \left(1 - \frac{E_r}{E_\nu} \right)^2 - g_L g_R \frac{m_e E_r}{E_\nu^2}\right] \, ,
\label{eq:xsec_nue}
\end{equation}
where we assume the free-electron approximation and $m_e$ is the electron mass.   
Here, the left-handed, $g_L = \left( g_V + g_A \right)/2$, and right-handed couplings,  $g_R = \left( g_V - g_A \right)/2$, are expressed in terms of the vector and axial vector couplings with
\begin{equation}
\begin{aligned}
g_V=& -1/2 + 2  \sin^2 \theta_W + \delta_{\alpha e} \,, \\
g_A=& - 1/2 + \delta_{\alpha e} \, .
\end{aligned}
\label{eq:nue-couplings}
\end{equation}
The cross-section for antineutrino scattering off electrons is obtained by exchanging $g_L \leftrightarrow g_R$. 
Notice that the factor $\delta_{\alpha e}$ in Eq.~(\ref{eq:nue-couplings}) is present only when electron (anti)neutrinos are involved.
In this case, the cross-section receives contributions from both neutral-current and charged-current interactions, unlike the case of $\alpha=\mu, \tau$, where the interaction is purely neutral-current.

As  we mentioned, Eq.~(\ref{eq:xsec_nue}) applies to \eves\ with free electrons.
To take into account electron binding effects in the target material of a given experiment, we weight the free \eves\ cross-section by adopting the step
approximation~\cite{Chen:2016eab}~\footnote{Note that we normalize the suppression factor to unity.} 
\begin{equation}
\left(\frac{\d \sigma_{\nu_\alpha-e^-}}{\d E_r}\right)_\text{SM} = \frac{1}{Z}\sum\limits_{i=1}^{Z} \Theta(E_r-B_i) \left(\frac{\d \sigma_{\nu_\alpha-e^-}}{\d E_r}\right)_\text{SM}^\text{free} \, ,
\label{eq:xsec-stepping}
\end{equation}
where $B_i$ is the binding energy of the $i$th atomic (sub)shell.
This way, one suppresses the free \eves cross-section and quantifies the impact of the atomic ionization energy levels. 
This calculation takes into account only those electrons that can be ionized by an energy deposition $E_r$, the modifications become important below a few keV recoil energies.

\subsection{Sterile neutrino dipole portal}
\label{sec:ster-neutr-dipole}

Recently, there has been some interest in the  transition of an active neutrino to a massive sterile state, induced by a magnetic coupling. 
Assuming a spin 1/2 nucleus, the corresponding \cevns\ cross-section reads~\footnote{Subdominant contributions due to a nuclear magnetic moment are neglected.}~\cite{McKeen:2010rx} 
\begin{equation}
\frac{\d\sigma_{\nu \mathcal{N} \rightarrow \nu_s \mathcal{N}}}{\d E_r} = \alpha_{\text{em}} \mu^2_{\nu,\text{eff}}  Z^2  \Bigg[\frac{1}{E_r}-\frac{1}{E_\nu}-\frac{m_{4}^2}{2E_{\nu}E_r M}\Bigg(1-\frac{E_r}{2E_\nu}+\frac{M}{2E_\nu}\Bigg)+\frac{m_{4}^4(E_r-M)}{8E_\nu^2 E_r^2 M^2}\Bigg] F_p^2(Q^2)\, ,
\label{eq:xsec_cevns_dipole}
\end{equation}
where $\alpha_{\text{em}}$ is the fine structure constant and $m_4$ is the sterile neutrino mass. For a spinless nucleus the differential cross-section remains essentially unchanged, i.e. 
\begin{equation}
\frac{\d \sigma_{(\text{spin}=1/2)}}{\d E_r} - \frac{\d \sigma_{(\text{spin}=0)}}{\d E_r} = \frac{m_4^2}{8 M E_\nu^2} \left(1+ \frac{m_4^2}{M E_r} \right) - \frac{E_r}{4 E_\nu^2} \, .
\end{equation}

For the case of the free-electron \eves\ via the neutrino dipole portal, the corresponding cross-section is trivially obtained from Eq.~(\ref{eq:xsec_cevns_dipole}) 
with the substitutions $M \to m_e$ and $Z^2 F_p^2(Q^2) \to 1 $, as~\cite{Shoemaker:2020kji,Brdar:2020quo} 
\begin{equation}
\left(\frac{\d\sigma_{\nu e^- \rightarrow \nu_s e^-}}{\d E_r}\right)^\text{free} =\alpha_{\text{em}}   \mu^2_{\nu,\text{eff}}  \Bigg[\frac{1}{E_r}-\frac{1}{E_\nu}-\frac{m_{4}^2}{2E_{\nu}E_r m_e}\Bigg(1-\frac{E_r}{2E_\nu}+\frac{m_e}{2E_\nu}\Bigg)+\frac{m_{4}^4(E_r-m_e)}{8E_\nu^2E_r^2m_e^2}\Bigg] \, .
\label{eq:xsec_nue_dipole}
\end{equation}
As in the SM case, the \eves\ cross-section is weighted with the step function in Eq.~(\ref{eq:xsec-stepping}). 
Note that for both, \cevns\ and \eves, we recover the usual expressions for the conventional neutrino magnetic moment cross-section as $m_4$ approaches zero~\cite{Vogel:1989iv}. 

Moreover, for the case of massive final state neutrinos, one might consider the interference term between magnetic and weak interactions. 
Neglecting the incident neutrino mass, the corresponding cross-section for $\bar{\nu}_e-e^{-}$ scattering can been written as~\cite{Grimus:1997aa} 
\begin{equation}
\left(\frac{\d \sigma_{\bar{\nu}_e e^-\to \nu_s e^- }}{\d E_r}\right)^\text{interf} = \frac{\alpha_\text{em} G_F \, m_4}{\sqrt{2} E_\nu \, m_e} \text{Re} \Bigg[ \sum_{j,n} e^{-i \frac{\Delta m^2_{jn} L}{2 E_\nu}} U_{ej} U^*_{en}  \,  \tilde{\lambda}_{j4}\left( \frac{m_e}{E_\nu} - \frac{E_r}{E_\nu} \right) Z^{V*}_{n4} + \left( 2 - \frac{E_r}{E_\nu}\right) Z^{A*}_{n4} \Bigg] \ ,
\end{equation}
where $Z^{V,A}_{jk} = U_{ej} U^*_{ek} + \delta_{jk} \tilde{g}_{V,A}$ with $\tilde{g}_V = -1/2 + 2 \sin^2 \theta_W$ and $\tilde{g}_A = -1/2$~\footnote{For $\nu_e-e^-$ scattering, the replacement $\tilde{g}_A \to - \tilde{g}_A$ and the appropriate complex conjugation of the mixing matrix elements should be made.}. 
From the latter expression it can be deduced that the cross-section is proportional to $\frac{m_4}{m_e} \sin 2 \theta_{14}$ for incident $\nu_e$ or $\bar{\nu}_e$.
Similarly, for incident $\nu_\mu$ or $\bar{\nu}_\mu$, the cross-section is proportional to $\frac{m_4}{m_e} c_{14} s_{24}^2$. 
For the case of CE$\nu$NS,  the interference cross-section can be obtained via the substitutions:
$\tilde{\lambda}_{ij} \to \tilde{\lambda}_{ij} Z F_p(q^2)$, $m_e \to M$, $\tilde{g}_V \to \mathcal{Q}_V$ and $\tilde{g}_A \to \mathcal{Q}_A$~\footnote{The axial weak charge $\mathcal{Q}_A$ vanishes for spin-zero nuclei and of the order of $\sim 1/A$ for nuclei with non-zero spin.}.

A few comments regarding the interference cross-section are in order.
First,  for sterile mixings of the order of $\sin^2\theta_{i4}\simeq 10^{-1}$, the interference effect will be tiny.  
  Moreover, due to its $1/E_\nu$ dependence, the cross-section is suppressed for neutrino energies above a few MeV, since from the kinematics one always has  $E_\nu$ larger than $m_4$.   
  Hence, experiments exposed to very low-energy neutrinos as the $^{51}$Cr source experiment we will discuss later, will be more sensitive to the interference effects than
  a reactor antineutrino experiment or a $\pi$-DAR based experiment, that will have negligible sensitivity.
  We finally note that the contribution from the CE$\nu$NS cross-section to this process can be safely ignored, as it is suppressed by the nuclear mass. 
  As an example, for the COHERENT experiment and $m_4=50$~MeV ($m_4=50$~keV), the interference-induced event rates are suppressed by seven (five) orders of magnitude compared to the purely magnetic-induced events.

\subsection{Event rates}
\label{sec:event-rates}   

In what follows, we simulate the expected signal for various \cevns\ and \eves\ experimental probes. The  differential event rate is written as
\begin{equation}
\frac{\d N}{\d E_r} = N_T \times \mathcal{E} \times \mathcal{A}(E_r) \int_{E_\nu^\text{min}}^{E_\nu^\text{max}} \frac{\d\sigma}{\d E_R}(E_\nu, E_r) \frac{\d \phi}{\d E_\nu}(E_\nu) \,\, \d E_\nu \, ,
\end{equation}
where $N_T$ denotes the number of targets (nuclei or electrons) per kg, $\mathcal{E}$ is the exposure in units of kg.yr and $\mathcal{A}(E_r)$ represents the efficiency. 
The differential cross-section is given by Eqs.~(\ref{eq:xsec-cevns})~and~({\ref{eq:xsec_nue}) for SM interactions only, or Eqs.~(\ref{eq:xsec_cevns_dipole})~and~(\ref{eq:xsec_nue_dipole})
  for the new physics scenario under consideration, while $\d \phi / \d E_\nu$ is the relevant neutrino flux for each experiment (see Sec.~\ref{sec:experiments}). 
  Finally, the upper integration limit is flux-dependent, while the lower integration limit is different for each process and given by 
\begin{equation}
\begin{aligned}
E_\nu^\text{min} = & \frac{1}{2}\left( E_r + \sqrt{E_r^2 + 2 M E_r} \right) \approx \sqrt{M E_r /2}, \qquad &\text{SM \cevns},\\
E_\nu^\text{min} \approx & \sqrt{M E_r /2} \left( 1 + \frac{m_4^2}{2 M E_r}\right), \qquad &\text{dipole portal \cevns}, \\
E_\nu^\text{min} = & \frac{1}{2}\left( E_r + \sqrt{E_r^2 + 2 m_e E_r} \right) , \qquad &\text{SM \eves},\\
E_\nu^\text{min} =& \frac{1}{2}\left(E_r + \sqrt{E_r^2 + 2 m_e E_r} \right) \left( 1 + \frac{m_4^2}{2 m_e E_r}\right), \qquad &\text{dipole portal \eves}.
\end{aligned}
\end{equation}
Notice that the limits corresponding to the sterile neutrino dipole interaction reduce to the SM ones in the limit of vanishing $m_4$.

\section{Experimental tests of neutrino magnetic moments}
\label{sec:experiments}

In this section, we explore the sensitivities of various present and future \cevns\ and \eves\ experiments to the sterile neutrino TMMs.
We present a brief discussion on each experimental facility,  describing all the necessary experimental inputs for our analysis.
 We must notice that, for the first three cases in this section, we describe in detail experiments that already have reported their data,
 while the last sub-section will be devoted to the future proposals that could show a better sensitivity to the neutrino magnetic moment. 

\subsection{XENON1T}
\label{sec:xenon1t}

The low-energy electron recoil data sample recorded with the XENON1T experiment~\cite{Aprile:2020tmw} has prompted a plethora of works attempting to interpret the anomaly.
 In particular, these data have been discussed within the framework of flavor transition magnetic moments in Ref.~\cite{Miranda:2020kwy}.
In the present paper we consider the sterile dipole portal, as discussed in~\cite{Shoemaker:2020kji,Brdar:2020quo,Karmakar:2020rbi}. 
Before addressing the various novel experimental setups of the coming subsections and their physics possibilities, we first calibrate our procedures with previous studies on XENON1T.
Besides confirming these earlier results, here we include the step function correction commented in Eq.~(\ref{eq:xsec-stepping}) and previously ignored in the literature,
since its impact is particularly relevant in the 1--7 keV region of the XENON1T excess.
Although the most relevant solar neutrino fluxes to consider are the $pp$ and the $^{7}\text{Be}$ neutrinos, in our calculation we include the full solar neutrino
spectrum from Ref.~\cite{Baxter:2021pqo}}. 
For the reconstructed energy $E$, we assume a Gaussian resolution function with $\sigma= a \sqrt{E} + b E$ and $a = 0.310 \sqrt{\mathrm{keV}}$ and $b = 0.0037~\mathrm{keV}$~\cite{Aprile:2020tmw}.
We also take into account the experimental exposure and efficiency provided by the XENON1T collaboration~\cite{Aprile:2020tmw}. 
  For the statistical analysis, we adopt the $\chi^2$ function~\cite{Boehm:2020ltd}
\begin{equation}
\chi^2 (\mathcal{S}) =  \sum_{i=1}^{29} \left(\frac{N_i^\text{exp} - N_i^\text{theor}(\mathcal{S})}{\sigma_i} \right)^2 + \left( \frac{\mathtt{a_1}-1}{0.03} \right)^2 + \left( \frac{\mathtt{a_2}-1}{0.026} \right)^2 \, ,
\end{equation}
where $N_i^\text{exp}$ is the observed number of events and $N_i^\text{theor}$ is the theoretical number of new physics events including backgrounds $B_0$. 
We include all the experimental data and errors from the data release in Ref.~\cite{Aprile:2020tmw}.
We also allow the overall normalization of the background ($\mathtt{a_2}$) and the efficiency ($\mathtt{a_1}$) to float with a Gaussian error of 2.6\%  and 3\%, respectively.

The above XENON1T exercise  has motivated us to explore the sterile neutrino dipole portal scenario within a broader context, using currently available low-energy \cevns\ and \eves\ data, to which we now turn.

\subsection{COHERENT}
\label{sec:coherent}   

The COHERENT collaboration has reported the \cevns\ detection in CsI~\cite{Akimov:2017ade} and liquid argon (LAr)~\cite{Akimov:2020pdx} detectors.
In this case, for the incoming neutrino flux, we consider the Michel spectrum, which describes the $\nu_e, \nu_\mu$, and $\bar{\nu}_\mu$ energy spectra
generated from pion decay at rest ($\pi$-DAR) for the Spallation Neutron Source (SNS)~\cite{Louis:2009zza}. 
The $\pi$-DAR flux is normalized to $\eta = r N_{\mathrm{POT}}/4 \pi L^2$, where $L$ is the baseline, while $N_{\mathrm{POT}}$ and $r$ denote the number of protons on target (POT)
and the number of neutrinos per flavor per POT, respectively~\footnote{The values for COHERENT-CsI and CENNS-10 detectors are taken from Refs.~\cite{Akimov:2017ade,Akimov:2020pdx}.}. 
For the 14.57~kg CsI detector, we calculate the theoretical signal as a function of the nuclear recoil energy in $\mathrm{keV_{nr}}$.
To convert this signal into the electron-equivalent energy space, $\mathrm{keV_{ee}}$, we use the new energy-dependent quenching factor, recently reported in Ref.~\cite{Konovalov:M7s}
(for its impact on  physics beyond the SM, see Ref.~\cite{Papoulias:2019txv}).  Finally, using the light yield $\text{LY}=13.348~\mathrm{PE/keV_{ee}}$~\cite{Akimov:2018vzs},
the signal is converted into a photoelectron (PE) spectrum, which we compare to the experimental data. 
We proceed in an analogous way for the case of the 24~kg CENNS-10 detector subsystem of COHERENT (hereafter COHERENT-LAr). 
We first evaluate the expected signal in $\mathrm{keV_{ee}}$ using the reported quenching factor, $\text{QF} = 0.246 + 7.8 \times 10^{-4}~\mathrm{keV_{nr}^{-1}}~E_r$.
Afterwards, following Ref.~\cite{Akimov:2020czh}, the signal is converted into the reconstructed energy, $E$,
using a normalized Gaussian function with resolution power $\sigma/E = 0.58/\sqrt{E(\mathrm{keV_{ee}})}$. 

For our statistical analysis, we consider the experimental \cevns\ data from the COHERENT-CsI and COHERENT-LAr measurements.
Concerning the CsI detector, we base our statistical analysis  on the $\chi^2$ function~\cite{Akimov:2017ade} 
\begin{widetext}
\begin{equation}
\begin{aligned}
\chi^2 (\mathcal{S}) =   \sum_{i=4}^{15} \left(\frac{N_i^{\mathrm{exp}} - N_i^{\mathrm{CE\nu NS}}(\mathcal{S}) [1+\mathtt{a}_1] - B^i_{0n} [1+\mathtt{a}_2] }{\sqrt{N_i^{\mathrm{exp}} + B_i^{0n} + 2 B_i^{ss}}}\right)^2 
  + \left(\frac{\mathtt{a}_1}{\sigma_{\mathtt{a}_1}} \right)^2 + \left(\frac{\mathtt{a}_2}{\sigma_{\mathtt{a}_2}} \right)^2  \, ,
\end{aligned}
\label{eq:chi}
\end{equation}
\end{widetext}
with $N_i^{\mathrm{exp}}$ and $N_i^{\mathrm{CE\nu NS}}$ being the measured and theoretical signal for the $i$th bin, respectively. 
The analysis is restricted to the 12 bins from $i=4$ to $i=15$, corresponding to $6 \leq \mathrm{PE} \leq 30$. 
$B_i^{0n}$ is the beam-on prompt neutron background, while $B^{ss}_i$ denotes the steady-state background events taken from the AC-ON data~\cite{Akimov:2018vzs}.
As explained in Refs.~\cite{Papoulias:2019txv, Akimov:2017ade}, the nuisance parameters $\mathtt{a}_1$ and $\mathtt{a}_2$ quantify the systematic
uncertainties of the signal and background rate, respectively, with $\sigma_{\mathtt{a}_1} = 12.8\%$ and $\sigma_{\mathtt{a}_2} = 25\%$. 

Concerning the COHERENT-LAr data, we focus on the \emph{analysis-A} of COHERENT~\cite{Akimov:2020pdx}, with our sensitivity analysis based on the $\chi^2$ function
defined in Ref.~\cite{Cadeddu:2020lky}
\begin{eqnarray}
\chi^2 (\mathcal{S})
&= &
\sum_{i=1}^{12}
\dfrac{\left(
N_{i}^{\text{exp}}
-
\eta_{\mathrm{CE\nu NS}} N_i^{\mathrm{CE\nu NS}}(\mathcal{S})
-
\eta_{\mathrm{PBRN}} B_i^{\mathrm{PBRN}}
-
\eta_{\mathrm{LBRN}} B_i^{\mathrm{LBRN}}\right)^2}
{\left( \sigma_i^{\mathrm{exp}} \right)^2 + \left[ \sigma_{\mathrm{BRNES}} \left( B_i^{\mathrm{PBRN}} + B_i^{\mathrm{LBRN}}\right)\right]^2}
\\ \nonumber
&+&
\left( \dfrac{\eta_{\mathrm{CE\nu NS}}-1}{\sigma_{\mathrm{CE\nu NS}}} \right)^2
+
\left( \dfrac{\eta_{\mathrm{PBRN}}-1}{\sigma_{\mathrm{PBRN}}} \right)^2
+
\left( \dfrac{\eta_{\mathrm{LBRN}}-1}{\sigma_{\mathrm{LBRN}}} \right)^2 .
\label{chi-spectrum}
\end{eqnarray}
Here we consider 12 bins in the range $\left[0,120 \right]~\mathrm{keV_{ee}}$ of the reconstructed energy, with $10~\mathrm{keV_{ee}}$ size each.
$N_{i}^{\text{exp}}$ denotes the measured signal with uncertainty $\sigma_i^\text{exp}$, BRNES corresponds to the Beam Related Neutron Energy Shape, while PBRN and LBRN stand for the
Prompt and Late Beam-Related Neutron Background data with $\sigma_{\mathrm{PBRN}}=32\%$ and  $\sigma_{\mathrm{LBRN}}=100\%$, respectively, taken from Ref.~\cite{Akimov:2020czh}.
The Beam Related Neutron Energy Shape (BRNES) uncertainty $\sigma_\text{BRNES}$ ($1.7\%$) and the systematic uncertainty of the signal rate $\sigma_{\mathrm{CE\nu NS}}$ ($13.4\%$) are taken from Ref.~\cite{Cadeddu:2020lky}.

\subsection{TEXONO} 
\label{sec:texono}

The TEXONO collaboration has reported a measurement of elastic neutrino-electron scattering (\eves) using a 187~kg CsI(Tl) detector at the Kuo-Sheng Nuclear reactor~\cite{Deniz:2009mu}. 
In this case, we consider the reactor antineutrino distribution from~\cite{Mention:2011rk},
normalized to  a total neutrino flux of $6.4 \times 10^{12}~\mathrm{cm^{-2} s^{-1}}$.
For energies below 2 MeV we adopt the theoretical estimations from Ref.~\cite{Kopeikin:1997ve}.
Our theoretical neutrino signal expected at the detector is expressed in units  of events/(kg $\cdot$ day $\cdot$ MeV) and compared with the experimental data.
Our statistical analysis follows from the $\chi^2$ function 
\begin{widetext}
\begin{equation}
\begin{aligned}
\chi^2 (\mathcal{S}) =  \sum_{i=1}^{10} \left(\frac{N_i^{\mathrm{meas}} - N_i^{\mathrm{new}}(\mathcal{S}) [1+\mathtt{a}] }{\sigma^\text{stat}_i }\right)^2   + \left(\frac{\mathtt{a}}{\sigma_\text{sys}} \right)^2    \, ,
\end{aligned}
\label{eq:chi2texono}
\end{equation}
\end{widetext}
with $N_i^\text{meas}$ standing for the detected events. We consider the reported 10 bins distributed over the recoil energy range
$[3,8]$~MeV as well as their associated  statistical errors,  $\sigma^\text{stat}_i$, as reported in Ref.~\cite{Deniz:2009mu}.
Systematic uncertainties are introduced through a penalty term  with $\sigma_\text{sys}=20\%$. As  the first step  in our procedure, we reproduce the TEXONO limit on the
effective electron antineutrino reactor magnetic moment, $\mu_{\bar{\nu}_e} \leq 2.2\times 10^{-10}~\mu_B$, ensuring that our statistical analysis is well-calibrated with TEXONO collaboration.

\subsection{\cevns\ and \eves\ with a $^{51}$Cr source}
\label{sec:cevns-eves-with}  

We are now motivated to explore the prospects of probing new physics phenomena using the intense beam of a low-energy monochromatic $\nu_e$  $^{51}$Cr source through either \cevns\ or E$\nu$ES processes.
Indeed, the very low detection thresholds aimed at these facilities, make them ideal for neutrino magnetic moment searches.
We first focus on a recent proposal for measuring \cevns\ using a 5 MCi $^{51}$Cr source and various kg-scale detectors with sub-keV capabilities~\cite{Bellenghi:2019vtc}.  
In accordance with the proposal, we consider a cylindrical $\mathrm{2000~cm^3}$ detector for different choices of target material such as Si, Ge, sapphire ($\mathrm{Al_2 O_3}$)
and calcium tungstate ($\mathrm{CaWO_4}$). The detector will be placed  $25$~cm from the source, what would imply an average neutrino flux of $1.1 \times 10^{13}~\mathrm{cm^{-2}  s^{-1}}$.
We estimate a target mass of $4.66$, $10.6$, $7.96$, and $12.12$~kg for Si, Ge, $\mathrm{Al_2 O_3}$, and $\mathrm{CaWO_4}$, respectively. 
A fixed threshold of $E_r^\text{thres} = 8~\mathrm{eV_{nr}}$ is considered. In our simulation, we include separately the contributions from the four lines with neutrino energies
(427, 432, 747, 752) keV that come from the $^{51}$Cr decay. Their relative strength is (9, 1, 81, 9)\%. The exposure time is expected to be $2$ half-lives i.e., $55.4$~days~\cite{Bellenghi:2019vtc}. 

In a previous work~\cite{Miranda:2020zji}, we have examined the potential of a $^{51}$Cr-LXe detector in probing new neutral gauge bosons through \eves\ measurements. 
We considered three different experimental configurations described in Ref.~\cite{Link:2019pbm}, i.e. (A, B, C): with initial radioactivity $R_{\mathrm{Cr} 51}^{0}= (5, 5, 10)$ MCi
and a time interval of $(100, 50, 50)$ days, respectively. 
In this work, we employ the same configuration and a cylindrical LXe detector, located at a distance of 1~m from the source, with a height and diameter of $1.38$~m,
which corresponds to a total mass of about 6 tonnes~\cite{Coloma:2014hka}.
The neutrino flux expected at a detector with this  geometry  is calculated as described in Ref.~\cite{Link:2019pbm}. 

Due to the lack of experimental data for estimating the sensitivity reach for this type of experiments, we employ a simplified statistical analysis based on the $\chi^2$ function
\begin{widetext}
\begin{equation}
\begin{aligned}
\chi^2 (\mathcal{S}) =  \sum_{i=1}^{n} \left(\frac{N_i^{\mathrm{SM}} - N_i^{\mathrm{new}}(\mathcal{S}) [1+\mathtt{a}] }{\sigma^\text{stat}_i }\right)^2   + \left(\frac{\mathtt{a}}{\sigma_\text{sys}} \right)^2    \, ,
\end{aligned}
\label{eq:chi-Cr51}
\end{equation}
\end{widetext}
with $\sigma^\text{stat}_i = \sqrt{N_i^\mathrm{SM} + N^\text{bg}_i}$. 
Here we have assumed a fixed background as large as 20\% of the SM rate, i.e. $N^\text{bg}_i = \sigma_\text{bg} N_i^\mathrm{SM}$ with $\sigma_\text{bg} = 20\%$,
while the systematic uncertainty is also taken to be $\sigma_\text{sys}=20\%$.
For the case of \cevns, we consider $n=12$ bins within the range $[E_r^\text{thres}, E_r^\text{max}]$ where $E_r^\text{max}$ is the maximum recoil energy for each nuclear target.
For \eves\ we take $5~\mathrm{keV_{ee}}$ wide bins in the range $[1,601]~\mathrm{keV_{ee}}$ (see Ref.~\cite{Link:2019pbm}).


\section{Results} 
\label{sec:results}
\begin{figure}[t]
\includegraphics[width=0.49\textwidth]{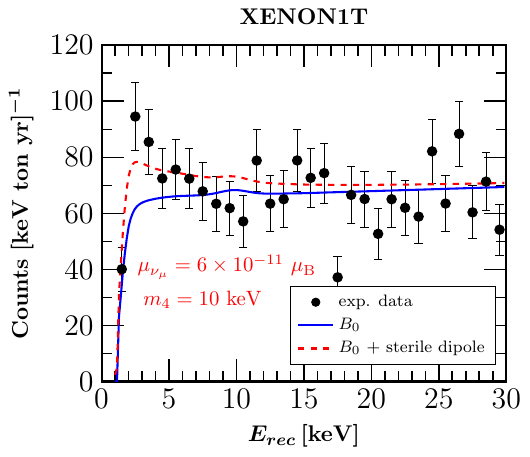}
\includegraphics[width=0.49\textwidth]{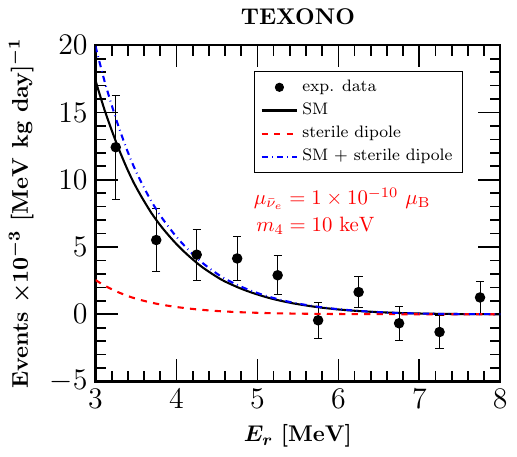}
\includegraphics[width=0.49\textwidth]{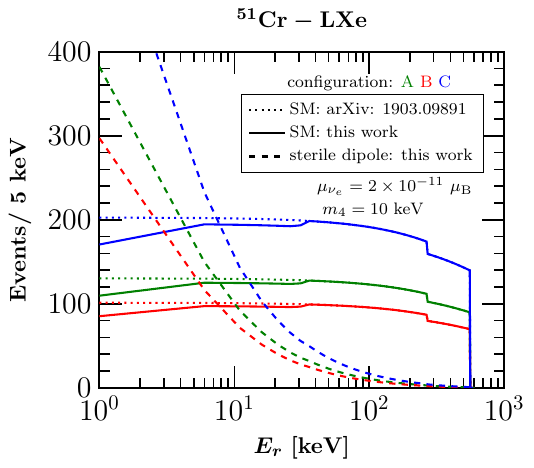}
\caption{Signal at XENON1T \textbf{(top left)} assuming  $\mu_{\nu_\mu} = 6 \times 10^{-11} \,  \mathrm{\mu_B}$ $^{51}$, TEXONO \textbf{(top right)}  assuming $\mu_{\bar{\nu}_e} = 1 \times 10^{-10} \,  \mathrm{\mu_B}$ and Cr-LXe \textbf{(bottom)} assuming $\mu_{\nu_e} = 2 \times 10^{-11} \,  \mathrm{\mu_B}$. All cases correspond to  a sterile neutrino with mass $m_4$=10 keV.}
\label{fig:events_nue}
\end{figure}

\begin{figure}[t!]
\includegraphics[width=\textwidth]{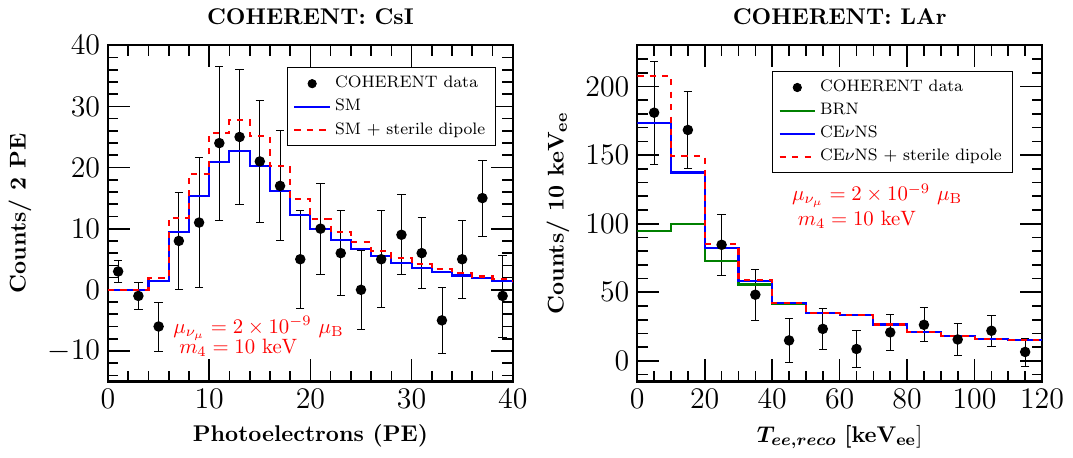}
\includegraphics[width=0.49\textwidth]{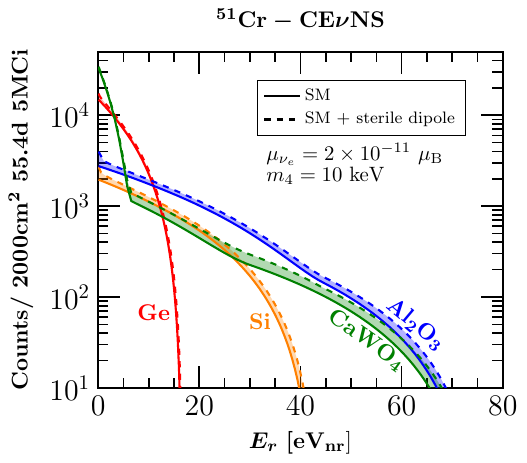}
\caption{Signal at COHERENT-CsI detector \textbf{(top left)}, COHERENT-LAr detector \textbf{(top right)} assuming a transition magnetic moment $\mu_{\nu_\mu} = 2 \times 10^{-9} \,  \mathrm{\mu_B}$ and at a $^{51}$Cr \cevns\ experiment \textbf{(bottom}) assuming a transition magnetic moment $\mu_{\nu_e} = 2 \times 10^{-11} \,  \mathrm{\mu_B}$. All cases correspond to  a sterile neutrino with mass $m_4$=10 keV.}
\label{fig:events_CEvNS}
\end{figure}

Here we present the results of the analysis described in the previous section. We have studied the sensitivity of current and future \cevns\ and \eves\ experiments
to the so-called dipole neutrino portal. 

Focusing first on the \eves\ experiments, we concentrate on the solar, reactor, and $^{51}$Cr neutrino fluxes, relevant for the XENON1T, TEXONO, and $^{51}$Cr-LXe experiments, respectively. 
For the aforementioned experiments, we illustrate in Fig.~\ref{fig:events_nue} the effect of a neutrino dipole moment assuming the production of a 10~keV sterile neutrino. 
In particular, the recent XENON1T excess is shown in the top left panel of Fig.~\ref{fig:events_nue}, where we present our results assuming the case of a $\nu_\mu$
(see Ref.~\cite{Shoemaker:2020kji} for the $\nu_\tau$ coupling) with the indicated benchmark value of effective neutrino magnetic moment and sterile neutrino mass.
Although there are small differences with the corresponding results of Ref.~\cite{Brdar:2020quo}, they are understandable because the authors did not consider the effect of the step function. 
Still, their analysis is consistent with the present one, because most of the corrections from the step function are washed out by the XENON1T efficiency. 
Since the effect of the step function is more pronounced for low threshold experiments, one should stress its importance for studies involving spectral features at low-energy recoils. 
In the top right panel in Fig.~\ref{fig:events_nue} we show the case of TEXONO, where the reported data points are plotted and compared with the sterile neutrino dipole moment expected signal. 
Although the effect is visible for these parameters, the statistical uncertainties are still large in this kind of measurements. 
Finally, in the bottom panel of Fig.~\ref{fig:events_nue} we show the number of events expected in the SM,
as well as in the sterile dipole portal, assuming the three different configurations (A, B, C) of the $^{51}$Cr-LXe detector described previously. 
Comparing our SM prediction with other reported studies, such as the one in Ref.~\cite{Link:2019pbm}, we find they essentially agree, except for the extra corrections associated to bound electrons
incorporated in our present work. 
\begin{figure}
\includegraphics[width=0.95\textwidth]{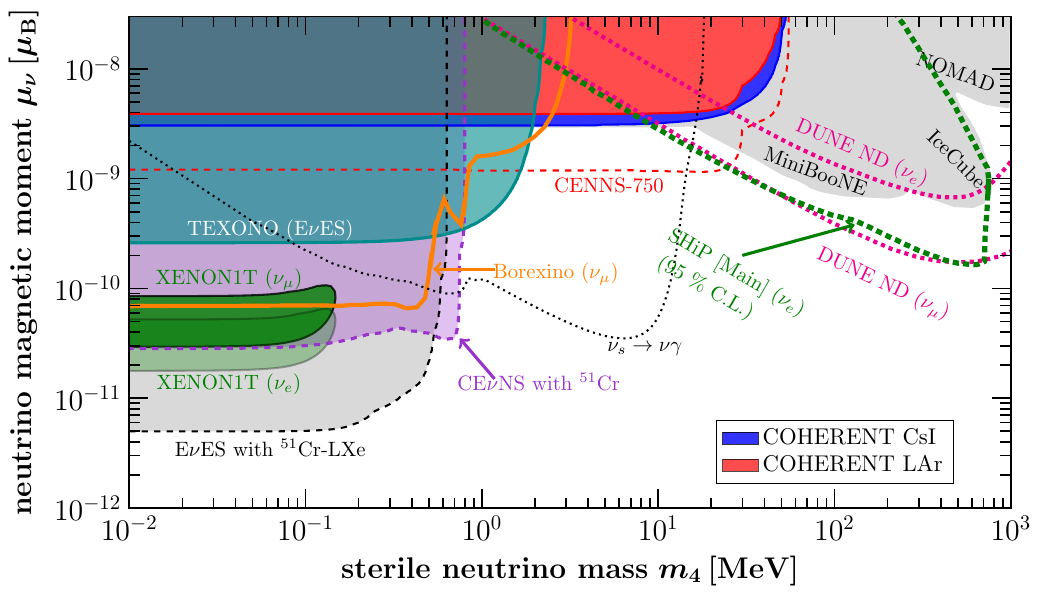}
\caption{Sensitivity of \cevns\ and \eves\ experiments to the effective sterile neutrino transition magnetic moments.
  Relevant limits from other experiments are shown for comparison. Solid (dashed) lines correspond to current (future) experiments (see text).}
\label{fig:contour}
\end{figure}
\begin{figure}
\includegraphics[width=0.95\textwidth]{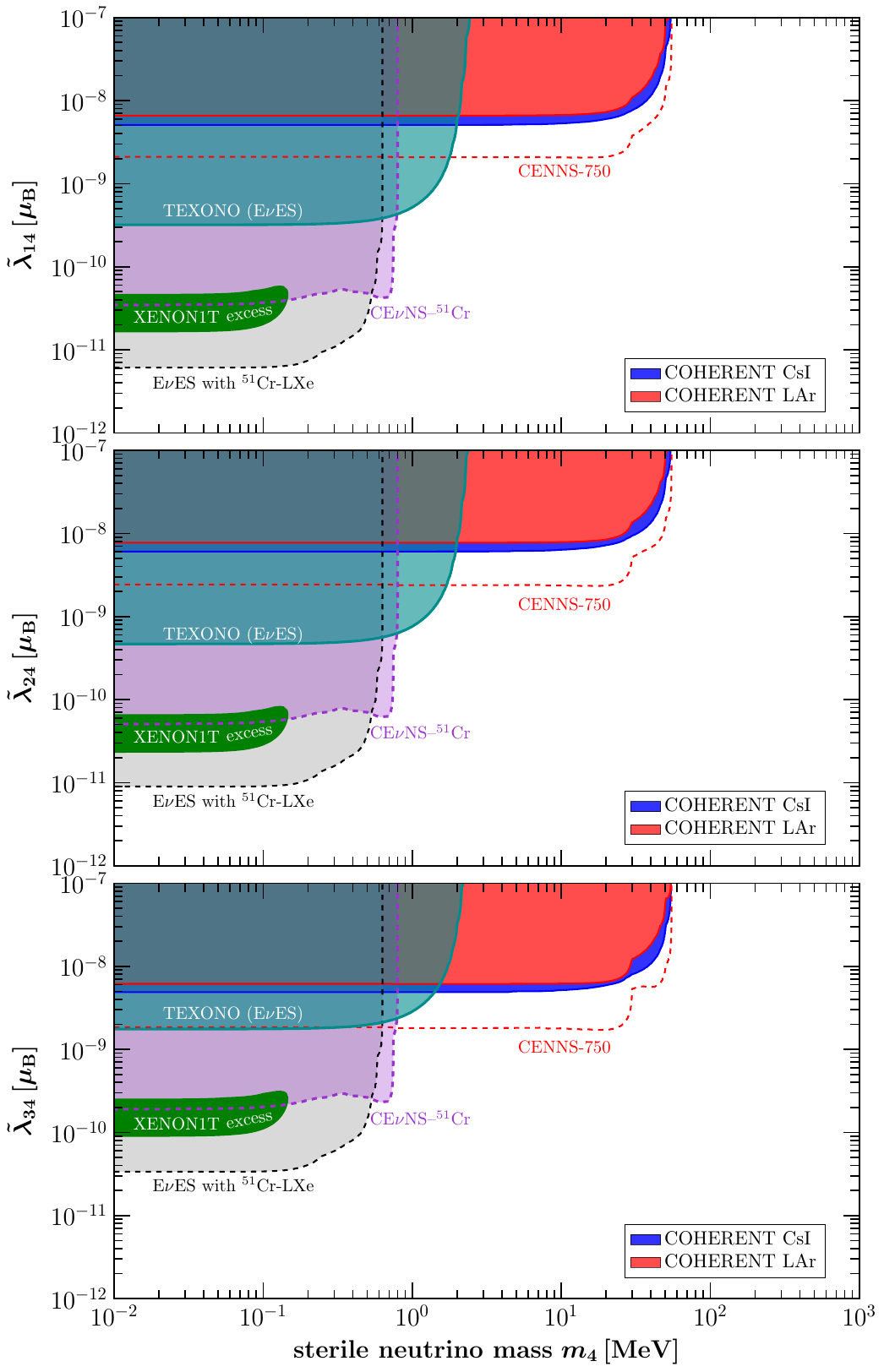}
\caption{Sensitivity of \cevns\ and \eves\ experiments in terms of the general TMM description. Note that our TMM notation provides a common basis for describing all possible experimental setups, see text.}
\label{fig:contour_TMM}
\end{figure}

We turn now our attention to the CE$\nu$NS case. 
The upper panel of Fig.~\ref{fig:events_CEvNS} illustrates the expected signal assuming a magnetic moment of $\mu_{\nu_\mu}= 2 \times 10^{-9} \mu_B$ and a sterile neutrino mass $m_4=10$~keV at the CsI (left)
and LAr (right) detector subsystems of COHERENT~\footnote{This value is allowed by the recent analyses of COHERENT-CsI~\cite{Papoulias:2019txv} and COHERENT-LAr~\cite{Miranda:2020tif}.}.
In the lower panel, we present the corresponding result for the proposed $^{51}$Cr \cevns\ experiment described above, assuming $\mu_{\nu_e} = 2 \times 10^{-11} \mu_B$ and various detector materials.
Notice that, for the case of the proposed $^{51}$Cr-type experiment, we have used a neutrino magnetic moment strength that is two orders of magnitude lower compared to the assumed value for COHERENT.
This is possible thanks to the extremely low detection threshold achievable with these future detectors. 
However, a $^{51}$Cr neutrino source  experiment will be restricted to sub-MeV neutrino masses, while for the higher energy $\pi$-DAR neutrinos at COHERENT,
the sensitivity holds up to masses $m_4 \lesssim m_\mu /2\simeq 50$~MeV. 

As a first step, assuming the neutrino magnetic moment as an effective parameter, we perform a sensitivity analysis for all the \cevns\ and \eves\ experiments discussed above.
A summary of our results is depicted in Fig.~\ref{fig:contour}, where we show current limits and projected sensitivities at 90\% C.L. 
 One sees that, for this case, the allowed sterile dipole
  moment required to account for the XENON1T excess lies in the range $ 2 \times 10^{-11} \lesssim \mu_{\nu_\mu} \lesssim 8 \times 10^{-11} \mu_B$, for a sterile neutrino mass up to 150~keV.
  Note also that our analysis of the XENON1T excess data for the case of the $\nu_\mu \to \nu_s$ transition agrees well with Ref.~\cite{Brdar:2020quo} and is shown here for comparison.
   Also shown, is the corresponding result derived in this work for the  $\nu_e \to \nu_s$ transition.
    As expected, the latter is consistent with a lower effective magnetic moment due to higher statistics.
    Indeed, the low-energy solar neutrino flux arriving at the Earth, to which XENON1T is mostly sensitive, contains more $\nu_e$ compared to $\nu_\mu$.
    For the $\nu_\tau \to \nu_s$ transition, see Ref.~\cite{Shoemaker:2020kji}.

    Encouraged by these results, we now proceed with the analysis of \cevns\ and \eves\ data in novel experimental setups that could probe the above parameter space.
As can be seen in Fig.~\ref{fig:contour}, the COHERENT data rules out the region with $\mu_{\nu_\mu} \gtrsim 3 \times 10^{-9} \mu_B$ and $m_4 \lesssim 50~\mathrm{MeV}$,
with the CsI detector performing slightly better compared to LAr~\footnote{Only $\nu_\mu \to \nu_s$ transitions are assumed for the analysis of COHERENT data. 
  The new CsI data recently reported by the COHERENT collaboration~\cite{Akimov:2021dab} will modify these constraints.
  However, since the experimental threshold remains the same as in the previous run, and the TMM signal is governed mainly by the low-energy behavior, we expect that the improvement will be rather mild.}.
Concerning the potential of future dipole moment probes, we also show the sensitivity region for the next generation CENNS-750 detector with 610~kg fiducial mass and 3 years of data
aquisition~\cite{Akimov:2019xdj}~\footnote{The backgrounds are taken from the estimations of Ref.~\cite{Miranda:2020syh}.}.
For the case of E$\nu$ES, the current restrictions are coming from the TEXONO experiment, and one sees how $\mu_{\bar{\nu}_e} \gtrsim 3 \times 10^{-10} \mu_B$ is excluded,
i.e. the constraint is improved by one order of magnitude for the neutrino dipole moment coupling, compared to COHERENT results. 
However, there is a sharp sensitivity loss at $m_4 \sim 10$~MeV, due to the kinematic cut imposed by the low energy of reactor neutrinos.  

Finally, we now turn to the expected sensitivities for a $^{51}$Cr source neutrino flux. Both for the case of  \cevns\ and \eves\ measurements,
they would lead to a breakthrough sensitivity reach of a neutrino magnetic moment in the region $\mu_{\nu_e} \sim 10^{-12}~\mu_B$. 
As in the case of reactor neutrinos, the very low energy of the emitted neutrinos at the $^{51}$Cr decay leads to a loss of sensitivity
for $m_4 \gsim 750$~keV.
 We also note that the \eves\ $^{51}$Cr-LXe case was found to be the only experimental setup where the interference cross-section is non-negligible with respect to the purely magnetic one. 

Before closing this discussion, we wish to emphasize the complementarity of the bounds derived here from the analysis of \cevns\ and \eves\ experiments,
  with those from oscillation experiments~\footnote{Only limits from laboratory experiments are discussed. For astrophysical limits, see Ref.~\cite{Magill:2018jla}.}.
  For the case of  the $\nu_\mu \to \nu_s$ channel in COHERENT, complementary constraints and sensitivities come from MiniBooNE~\cite{MiniBooNE:2007uho}, NOMAD~\cite{Magill:2018jla},
  IceCube~\cite{Coloma:2017ppo},  Borexino~\cite{Brdar:2020quo} and DUNE near detector (ND)~\cite{Schwetz:2020xra,Atkinson:2021rnp}.
  One can see that, though not placing severe constraints on the neutrino magnetic moment, the current and future COHERENT data cover a large portion of the previously unexplored parameter space,
  overlapping with regions already probed by the aforementioned large scale experiments.
We should also mention that, although not visible, COHERENT-CsI is competitive to CHARM-II constraints~\cite{Coloma:2017ppo}, while the latter will be completely overridden by the future CENNS-750 experiment.
    Finally, also shown is the sensitivity obtained from  the $\nu_s \to \nu \gamma$ decay in Ref.~\cite{Plestid:2020vqf}, which is clearly complementary to the CENNS-750 and $^{51}$Cr \cevns\ experiments studied here.
Similarly, for the case of TEXONO and $^{51}$Cr-based \cevns\ and \eves\ experiments ($\nu_e \to \nu_s$ transitions), the relevant experiments would be XENON1T, SHiP and DUNE ND.
Using this one may compare our results in Fig.~\ref{fig:contour} for the proposed Chromium experiments with the relevant sensitivities from SHiP~\cite{Magill:2018jla} 
as well as DUNE ND~\cite{Schwetz:2020xra,Atkinson:2021rnp}. One sees that there is no overlap with the \cevns\ and \eves\ experiments.
In contrast, the $^{51}$Cr-LXe setup discussed here can provide an independent test of the region indicated by the XENON1T excess~\footnote{Bounds from existing nuclear recoil XENON1T data~\cite{Shoemaker:2018vii} and future LHC projections~\cite{Ismail:2021dyp} are weaker.}.
However, we emphasize that one should really use Eq.~(\ref{eq:nmm_solar}) for the effective magnetic moment for solar neutrino experiments. Only such general $\lambda$-formalism provides a basis for making such comparisons.
  
Having presented our results for effective neutrino magnetic moments, we now explore the current and future sensitivities of the aforementioned \cevns\ and \eves\ experiments for the TMMs, as
expressed within the general formalism discussed in Sec.~\ref{sec:basic-formalism}. 
As emphasized at the Introduction, in contrast to the simple effective magnetic moment description, the adoption of the more general TMMs formalism allows for a direct
comparison of the attainable sensitivities at different types of experiments in terms of the same fundamental parameters, $\tilde{\lambda}_{ij}$.
Moreover, adopting this general formalism allows the full data set of experiments exposed to a neutrino source with multiple flavors to be used in combined analyses.
Hence, for COHERENT and XENON1T, we do not need to consider one non-zero effective magnetic moment $\mu_{\nu_\alpha}$ ($\alpha=e,\mu,\tau$) at a time.
In what follows, all relevant $\mu_{\nu_\alpha}$ will be assumed non-vanishing and will be expressed in terms of the basic TMMs $\lambda_{ij}$.
We present our results for a simplified case, assuming only one non-vanishing TMM $\tilde{\lambda}_{ij}$ at a time, and neglecting the associated CP phases
(for a discussion on the impact of the CP phases see Ref.~\cite{Miranda:2019wdy}).
The current constraints and future sensitivities are shown in Fig.~\ref{fig:contour_TMM}, where one sees the same qualitative behavior as in the case of effective dipole moments.  

\section{Conclusions and outlook}
\label{sec:conclusions-outlook}

Motivated by the XENON1T excess and by the intrinsic interest in probing neutrino electromagnetic properties, we have examined the current and future experimental sensitivities to
a dipole portal interaction associated to a massive sterile neutrino with transition magnetic moment.
We have explored such scenario for various \cevns\ and \eves\ experimental setups, analyzing their potential in probing the region of interest for the XENON1T excess.

Besides presenting the relevant sensitivities in terms of the usual effective magnetic moments, we have given the first comprehensive description in terms of the fundamental TMM parameters, see Tables~\ref{tab:coefficients_nue} and \ref{tab:coefficients_numu}. 
Interference between weak and magnetic terms, possible for massive sterile neutrinos, has been found to play no essential role in constraining the parameters.
The only exception to this was found when considering very-low energy \eves\ from a  $^{51}$Cr source.  

Our phenomenological analysis has focused on current and future \cevns\ and \eves\  experiments using low-energy neutrinos from artificial neutrino sources, such as reactors and accelerators, as well as those emerging from a $^{51}$Cr source.
Our analysis shows that the current constraints arising from the recent COHERENT \cevns\ measurements on CsI and LAr, as well as from reactor neutrino \eves\ measurements by TEXONO, can cover a wider, previously unexplored, region in sterile neutrino parameters.
In particular, we have shown that the proposed $^{51}$Cr experiments can fully probe the explanation of the XENON1T anomaly with the sterile dipole portal (see Figs.~\ref{fig:contour} and~\ref{fig:contour_TMM}).
Finally, we have also emphasized the complementarity of future low-energy \cevns\ and \eves\ experiments with large-scale experiments, such as DUNE ND, SHiP, Borexino, MiniBooNE, IceCube, and NOMAD, as seen in Sec.~\ref{sec:results}.

\acknowledgements 

\noindent 

  Work supported by the Spanish grants PID2020-113775GB-I00 (AEI / 10.13039/501100011033) and PROMETEO/2018/165 (Generalitat Valenciana),
  by Fundac\~ao para a Ci\^encia e a Tecnologia (FCT, Portugal) through grant CERN/FIS-PAR/0004/2019 and by
  CONACYT-Mexico under grant A1-S-23238. O. G. M. has been supported by SNI (Sistema Nacional de Investigadores).
The work of DKP is co-financed by Greece and the European Union (European Social Fund-ESF) through the Operational Programme ``Human Resources Development,
Education and Lifelong Learning'' in the context of the project ``Reinforcement of Postdoctoral Researchers - 2nd Cycle'' (MIS-5033021), implemented by the State Scholarships Foundation (IKY).

\providecommand{\href}[2]{#2}\begingroup\raggedright\endgroup


\end{document}